\documentclass[a4paper,11pt]{article}
\pdfoutput=1 

\usepackage{jheppub} 

\usepackage[T1]{fontenc} 
\usepackage{simpler-wick}	
\usepackage{mathrsfs}
\usepackage{amsmath}
\usepackage{hyperref}
\usepackage{tikz-feynman}
\usepackage{graphicx}
\usepackage{dcolumn}
\usepackage{bm}
\def\p{\mbox{\boldmath$\displaystyle\boldsymbol{p}$}}

\def\0{\mbox{\boldmath$\displaystyle\boldsymbol{0}$}}

\def\x{\mbox{\boldmath$\displaystyle\boldsymbol{x}$}}
\def\y{\mbox{\boldmath$\displaystyle\boldsymbol{y}$}}

\newcommand{\gdualn}[1]{\overset{\:{}^{{}^{\boldsymbol{\neg}}}}{\smash[t]{#1}}} 

\DeclareMathAlphabet{\mathsfit}{T1}{\sfdefault}{\mddefault}{\sldefault}
\SetMathAlphabet{\mathsfit}{bold}{T1}{\sfdefault}{\bfdefault}{\sldefault}

\makeatletter
\newsavebox{\@brx}
\newcommand{\llangle}[1][]{\savebox{\@brx}{\(\m@th{#1\langle}\)}%
  \mathopen{\copy\@brx\mkern2mu\kern-0.9\wd\@brx\usebox{\@brx}}}
\newcommand{\vrt}[1][]{\savebox{\@brx}{\(\m@th{#1\vert}\)}%
  \mathopen{\copy\@brx\mkern2mu\kern-0.9\wd\@brx\usebox{\@brx}}}  
\newcommand{\rrangle}[1][]{\savebox{\@brx}{\(\m@th{#1\rangle}\)}%
  \mathclose{\copy\@brx\mkern2mu\kern-0.9\wd\@brx\usebox{\@brx}}}
\makeatother

\title{Generalized unitary evolution for symplectic scalar fermions}

\author{Cheng-Yang Lee}

\affiliation{Center for Theoretical Physics, College of Physics,\\ Sichuan University, Chengdu 610064, China}

\emailAdd{cylee@scu.edu.cn}

\abstract{
The theory of symplectic scalar fermion of LeClair and Neubert is studied. The theory evades the conventional spin-statistics theorem because its Hamiltonian is pseudo Hermitian. The definition of pseudo Hermiticity is examined in the interacting and the Heisenberg picture. For states that evolve under pseudo Hermitian Hamiltonians, we define the appropriate inner-product and matrix element of operators that preserve time translation symmetry. The resulting $S$-matrix is shown to satisfy the generalized unitarity relation. We clarify the derivation of the symplectic currents and charges. By demanding the currents and charges to be pseudo Hermitian, the global symmetry of the free Lagrangian density reduces from $\text{Sp}(2,\mathbb{C})$ to $\text{SU}(2)$. By explicit calculations, we show that the LeClair-Neubert model of $N$ quartic self-interacting scalar fermions admits generalized unitary evolution. 
}

\begin{document} 
\maketitle
\flushbottom

\section{Introduction}
\label{sec:intro}

In quantum mechanics and quantum field theory, physical operators such as the Hamiltonian, are postulated to be Hermitian because their eigenvalues, which are observables, are real. Transformations generated by Hermitian operators admit unitary evolution of physical systems which preserve transition probabilities. The synthesis of unitarity and Lorentz symmetry inevitably lead to the celebrated spin-statistics theorem~\cite{Duck:1997ua,Streater:1989vi,Weinberg:1995mt}.


The postulates in quantum mechanics and the spin-statistics theorem have played indispensable roles in our ongoing efforts to understand the fundamental constituents of matter. And yet, according to the $\Lambda$CDM model and measurements from the Planck Collaboration, elementary particles of the Standard Model (SM) accounts only up to 5\% of the total energy and matter contents of the observed universe~\cite{Planck:2018vyg}. Therefore, it may be premature to suppose the present theoretical framework and their theorems are fixed. In fact, the development of physics from the beginning of the twentieth century has taught us that progresses are made by continual modifications and generalizations of the foundational axioms.\footnote{In essence, we are paraphrasing Dirac's view on the development of physics which remains very much relevant even in the twenty-first century~\cite{Dirac:1931kp}.} 

In this spirit, the works on $\mathcal{PT}$ symmetric Hamiltonians by Bender and Boettcher~\cite{Bender:1998ke,Bender:2007nj} and its subsequent generalization to pseudo Hermitian Hamiltonians by Mostafazadeh~\cite{Mostafazadeh:2001jk,Mostafazadeh:2008pw}, are of central importance. These works represent an expansive program to extend quantum mechanics and quantum field theory beyond the formalism of Hermitian operators. To be precise, a pseudo Hermitian Hamiltonian is defined by the following equation
\begin{equation}
H^{\#}\equiv\eta^{-1}H^{\dag}\eta=H,\label{eq:pseudo_H}
\end{equation}
where $\eta$ is an operator to be determined. While definition~(\ref{eq:pseudo_H}) was already known to Pauli~\cite{RevModPhys.15.175}, its physical implication was only realized after Mostafazadeh proved two important theorems concerning the spectrum of $H$ and the generalized unitary evolution of states when equipped with the appropriate inner-product. To facilitate ensuing discussions, we now present the two theorems of Mostafazadeh:
\begin{quote}
\textit{Theorem 1.} Let $|\alpha\rangle$ be an eigenstate of $H$ with eigenvalue $E_{\alpha}$. From~(\ref{eq:pseudo_H}), the eigenvalue $E_{\alpha}$ is real when $\eta_{\alpha}=\langle\alpha|\eta|\alpha\rangle\neq0$. 
\end{quote}
\begin{quote}
\textit{Theorem 2.} The invariant inner-product between two states is $\langle\beta|\alpha\rangle_{\eta}\equiv\langle\beta|\eta|\alpha\rangle$.
The time translations are the canonical quantum mechanical transformations, $|\alpha(t)\rangle= e^{-iHt}|\alpha\rangle$ and $\langle\beta(t)|=\langle\beta| e^{iH^{\dag}t}$, so that $\langle\beta(t)|\alpha(t)\rangle_{\eta}=\langle\beta|\alpha\rangle_{\eta}$. 
\end{quote}
Physically well-defined free quantum field theories with pseudo Hermitian adjoints that evade the spin-statistics theorem have been constructed in the spin-zero and spin-half representations~\cite{LeClair:2007iy,Ahluwalia:2022ttu,Ahluwalia:2022zrm,Ahluwalia:2022yvk,Ryu:2022ffg}.  The author and collaborators have, from first principle, constructed spin-half bosonic as well as fermionic fields of mass dimension one and three-half~\cite{Ahluwalia:2022ttu,Ahluwalia:2022zrm,Ahluwalia:2022yvk}. Despite the use of pseudo Hermitian adjoints, the free theories are unitary. That is, the free Hamiltonians are Hermitian and positive-definite. In the presence of interactions, the Hamiltonians become pseudo Hermitian so it is necessary to use the $\eta$ product to preserve time translation symmetry. An important problem for pseudo Hermitian theories is that the $\eta$ is not positive-definite so further works are needed to establish consistency for the interacting theories. For now, we are able to demonstrate that the $S$-matrix satisfy the generalized unitarity relation $S^{\#}S=I$, where $S^{\#}=\eta^{-1}S^{\dag}\eta$ but this does not reduce to $S^{\dag}S=I$. 


In this paper, we study the theory of symplectic scalar fermion constructed by LeClair and Neubert (LN)~\cite{LeClair:2007iy}. This theory is local, Lorentz-invariant, admits a positive-definite free Hamiltonian but furnishes fermionic statistics. The point of departure from the bosonic scalar field theory comes from the introduction of pseudo Hermitian adjoint. Due to the fermionic statistics, the theory is shown to have global symplectic symmetry and its $\beta$ function, in the case of quartic self-interaction has non-trivial fix point~\cite{LeClair:2007iy}. The theory has found applications in conformal field theory~\cite{Ryu:2022ffg} and dS/CFT correspondence~\cite{Anninos:2011ui,Ng:2012xp,Das:2012dt,Sato:2015tta,Narayan:2016xwq,Fei:2015kta}. For the model of $N$ self-interacting scalar fermions (the LN model), we show that the 2 to 2 scattering process is unitary below the pair-production threshold, so the model may have applications in condensed matter physics.

The paper is organized as follows. In sec.~\ref{scalar_fermion}, we review the theory of scalar fermion and the properties of pseudo Hermitian Hamiltonians. In sec.\ref{sym}, we clarify the derivation of the global symplectic symmetry. In sec.~\ref{opt}, we show that for the LN model, the $S$-matrix satisfies the generalized unitarity relation.

\section{Scalar fermions}\label{scalar_fermion}


Let $\phi$ be a complex scalar field and $\gdualn{\phi}$ be its adjoint. LeClair and Neubert made the crucial observation that $\gdualn{\phi}$ does not have to be the Hermitian conjugate of $\phi$. In fact, the following expansions
\begin{align}
\phi(x)&=(2\pi)^{-3/2}\int\frac{d^{3}p}{\sqrt{2E}}\left[e^{-ip\cdot x}a(\p)+e^{ip\cdot x}b^{\dag}(\p)\right]\label{eq:phi},\\
\gdualn{\phi}(x)&=(2\pi)^{-3/2}\int\frac{d^{3}p}{\sqrt{2E}}\left[e^{ip\cdot x}a^{\dag}(\p)-e^{-ip\cdot x}b(\p)\right],\label{eq:dphi}
\end{align}
are also legitimate. The demand of locality, and the minus sign in~(\ref{eq:dphi}) force $\phi$ and $\gdualn{\phi}$ to be fermionic rather than bosonic while respecting Lorentz symmetry. That is, at equal time, they anti-commute with each other $\{\phi(t,\x),\gdualn{\phi}(t,\y)\}=0$. In fact, they also anti-commute at the same space-time point $\{\phi(x),\gdualn{\phi}(x)\}=0$. Given an arbitrary Lorentz transformation $\Lambda$, they transform as
\begin{align}
U(\Lambda)\phi(x)U^{-1}(\Lambda)&=\phi(\Lambda x),\\
U(\Lambda)\gdualn{\phi}(x)U^{-1}(\Lambda)&=\gdualn{\phi}(\Lambda x),
\end{align}
where $U(\Lambda)$ is the unitary representation of $\Lambda$ in the Hilbert space. 
Taking into account of the fermionic statistics, the free propagator and Lagrangian density are given by
\begin{equation}
S(x-y)=\frac{i}{(2\pi)^{4}}\int d^{4}q\left[e^{-iq\cdot(x-y)}\frac{1}{q^{2}-m^{2}+i\epsilon}\right],
\end{equation}
and
\begin{equation}
\mathscr{L}=\partial^{\mu}\gdualn{\phi}\partial_{\mu}\phi-m^{2}\gdualn{\phi}\phi.\label{eq:L_phi}
\end{equation}
One can readily verify that the fields and their conjugate momenta satisfy the canonical equal-time anti-commutation relations and that the free Hamiltonian is positive-definite after normal-ordering~\cite{LeClair:2007iy}
\begin{equation}
H_{0}=\int d^{3}p\sqrt{|\p|^{2}+m^{2}}\left[a^{\dag}(\p)a(\p)+b^{\dag}(\p)b(\p)\right].
\end{equation} 
Because $\phi$ and $\gdualn{\phi}$ furnish fermionic statistics, we \textit{cannot} add a Hermitian conjugate term $\mathscr{L}^{\dag}$ to~(\ref{eq:L_phi}) to make the Lagrangian density Hermitian as it would lead to non-locality and non-unitarity because $\phi$ does not anti-commute with $\phi^{\dag}$. Similarly, given a pseudo Hermitian interacting potential, we cannot add a Hermitian conjugate to it. Therefore, the kinematics and dynamics of scalar fermions can only be described by $\gdualn{\phi}$ and $\phi$. Because $\gdualn{\phi}\neq\phi^{\dag}$, the free Lagrangian density is non-Hermitian. Nevertheless, the free propagator remains the scalar propagator and the free Hamiltonian is Hermitian. Therefore, the states have unitary evolution under $H_{0}$ and the fields evolve via
\begin{align}
\gdualn{\phi}(t,\x)=e^{iH_{0}t}\gdualn{\phi}(0,\x)e^{-iH_{0}t}, \\
\phi(t,\x)=e^{iH_{0}t}\phi(0,\x)e^{-iH_{0}t}.
\end{align}
Similarly, the free momentum operators are also Hermitian so the states and fields are unitary under spatial translation. 

But in the interacting picture, the full Hamiltonians that are functions of $\gdualn{\phi}\phi$ and its generalizations are non-Hermitian. To deal with non-Hermiticity, we first review and elaborate on the observation made by LN. As for how states evolve under such Hamiltonians, we defer the discussions to sec.~\ref{opt}. LeClair and Neubert noted that while $\gdualn{\phi}\phi$ is non-Hermitian, it is pseudo Hermitian~\cite{LeClair:2007iy}. There exists a Hermitian operator $\eta$ such that
\begin{equation}
\left[\gdualn{\phi}(0,\x)\phi(0,\x)\right]^{\#}\equiv\eta^{-1}\left[\gdualn{\phi}(0,\x)\phi(0,\x)\right]^{\dag}\eta=\gdualn{\phi}(0,\x)\phi(0,\x).\label{eq:phi_phi}
\end{equation}
Since $[H_{0},\eta]=O$,~(\ref{eq:phi_phi}) holds at all times under the evolution of $H_{0}$. We require $\eta$ to be Hermitian so that $(\gdualn{\phi}\phi)^{\#\#}=\gdualn{\phi}\phi$.
To find $\eta$, we take $\gdualn{\phi}$ to be the pseudo Hermitian conjugate of $\phi$ in the sense that that~\cite{LeClair:2007iy}
\begin{equation}
\gdualn{\phi}(x)=\eta^{-1}\phi^{\dag}(x)\eta. \label{eq:p_phi}
\end{equation}
Expanding~(\ref{eq:p_phi}) using~(\ref{eq:phi}-\ref{eq:dphi}) yields
\begin{align}
\eta^{-1}a(\p)\eta =a(\p),\quad
\eta^{-1}b^{\dag}(\p)\eta =-b^{\dag}(\p),\label{eq:eta_ab}
\end{align}
which are equivalent to
\begin{equation}
\eta^{\dag}|a(\p)\rangle=|a(\p)\rangle,\quad \eta|b(\p)\rangle=-|b(\p)\rangle,\label{eq:eta_ab2}
\end{equation}
where $a^{\dag}|0\rangle=|a\rangle$, $b^{\dag}|0\rangle=|b\rangle$
and we have demanded that the vacuum to be invariant under the action of $\eta$. We note, if the scalar field is taken to be real, then there is no non-trivial $\eta$. This tells us that a consistent theory of scalar fermion \textit{cannot} be formulated in terms of real scalar fields. Doing so inevitably leads to non-locality and non-unitarity.

Equation~(\ref{eq:eta_ab2}) can now be solved by the ansatz $\eta=\exp(i\theta\chi)$ where $\chi\equiv\int d^{3}p[b^{\dag}(\p)b(\p)]$ and $\theta\in\mathbb{R}$ is a phase to be determined. Acting $\eta$ on the particle and anti-particle state using the ansatz, we obtain $\eta^{\dag}|a\rangle=|a\rangle$ and $\eta|b\rangle=e^{i\theta}|b\rangle$. Choosing $\theta=-\pi$, we obtain~\cite{Robinson:2009xm}
\begin{equation}
\eta=\exp\left[-i\pi\int d^{3}p\, b^{\dag}(\p)b(\p)\right],\label{eq:eta}
\end{equation}
so the operator is unitary. Next, act $\eta$ on an arbitrary state $|\alpha\rangle$ successively, we obtain
\begin{equation}
\eta^{2}|\alpha\rangle=e^{-2in_{\alpha}\pi}|\alpha\rangle=|\alpha\rangle
\end{equation}
where $n_{\alpha}$ is the number of anti-particle states in $|\alpha\rangle$. Therefore, $\eta^{2}=\eta^{\dag}\eta=I$ so $\eta$ is Hermitian. Using~(\ref{eq:eta_ab}), we find $[H_{0},\eta]=O$. Therefore, the free Lagrangian density is pseudo Hermitian
\begin{equation}
\mathscr{L}^{\#}_{0}\equiv\eta^{-1}\mathscr{L}^{\dag}_{0}\eta=\mathscr{L}_{0}.
\end{equation}
Similarly, interactions that are functions of $\gdualn{\phi}\phi$ and its generalizations are pseudo Hermitian at all times. If we perform transformations on the fields
$\phi'=U\phi$, $\gdualn{\phi}'=\gdualn{\phi}U^{\#}$, we find that $\gdualn{\phi}'\phi'$ remains pseudo Hermitian provided that $U$ satisfies $U^{\#}U=I$.

\subsection{Pseudo Hermitian Hamiltonians}

We study the time evolutions of the scalar fermionic fields in the Heisenberg picture to demonstrate that the associated definition of pseudo Hermiticity is consistent with its counterpart in the interacting picture. Pseudo Hermitian Hamiltonian is defined as
\begin{equation}
H^{\#}\equiv\eta^{-1}H^{\dag}\eta=H\label{eq:eHe}
\end{equation}
where $[\eta,H]\neq O$ so that $H^{\dag}\neq H$. In the Heisenberg picture, this means that $\eta$ has a time dependence. Taking $\eta$ to be at time $t=0$, we obtain
\begin{equation}
\eta_{H}(T)\equiv e^{iHT}\eta e^{-iHT}\neq\eta.
\end{equation}
If we instead define pseudo Hermiticity using $\eta_{H}(T)$ where $T\neq0$, it is equivalent to~(\ref{eq:eHe}) since 
\begin{equation}
\eta^{-1}_{H}(T)H^{\dag}\eta_{H}(T)=\eta^{-1}H^{\dag}\eta=H.\label{eq:eHe_t}
\end{equation}
Equations~(\ref{eq:eHe}) and~(\ref{eq:eHe_t}) show that the time translation on $\eta$ induces a non-uniqueness in the definition of pseudo Hermiticity. Also, any similarity transformations $\eta\rightarrow\eta'=U\eta U^{-1}$ where $[U,H]=O$ satisfies~(\ref{eq:eHe}) and~(\ref{eq:eHe_t}). 

If we only consider the definition of pseudo Hermiticity of the Hamiltonian, then $\eta$ is unique up to similarity transformation and time translation. However, these considerations do not take into account of the fact that Hamiltonians are constructed from quantum fields. The demand of pseudo Hermiticity on quantum fields imposes additional  constraint on $\eta$. For scalar fermionic fields, the relation $\gdualn{\phi}=\eta^{-1}\phi^{\dag}\eta$ uniquely fix $\eta$ to be~(\ref{eq:eta_ab2}) thus removing the non-uniqueness. 

Now we consider pseudo Hermiticity in the Heisenberg picture. Let $\phi_{H}$ and $\gdualn{\phi}_{H}$ be the scalar fermionic fields in the Heisenberg picture. At $t=0$, we have
\begin{align}
\phi_{H}(0,\x)&=\phi(0,\x),\label{eq:phi_H} \\
\gdualn{\phi}_{H}(0,\x)&=\gdualn{\phi}(0,\x). \label{eq:gphi_H}
\end{align}
We take their time evolutions to be
\begin{align}
\phi_{H}(x)&\equiv e^{iHt}\phi(0,\x)e^{-iHt},\label{eq:phi_H_t}\\
\gdualn{\phi}_{H}(x)&\equiv e^{iHt}\gdualn{\phi}(0,\x)e^{-iHt}.\label{eq:dphi_H_t}
\end{align}
Because the Hamiltonian is pseudo Hermitian, the demand that both $\phi_{H}$ and $\gdualn{\phi}_{H}$ have the same time evolution requires further justification. Specifically, we need to demonstrate that no inconsistencies arise from~(\ref{eq:phi_H}-\ref{eq:dphi_H_t}). Towards this end, we rewrite~(\ref{eq:gphi_H}) as
$\phi^{\dag}=\eta\gdualn{\phi}\eta^{-1}$ and evolve $\phi^{\dag}$ to obtain
\begin{align}
\phi^{\dag}_{H}(x)&=e^{iH^{\dag}t}\phi^{\dag}(0,\x)e^{-iH^{\dag}t} \nonumber\\
&=e^{iH^{\dag}t}\left[\eta\gdualn{\phi}(0,\x)\eta^{-1}\right]e^{-iH^{\dag}t}.
\end{align}
Since $H^{\dag}\neq H$, the time evolution of $\phi^{\dag}_{H}$ is different from $\phi_{H}$. Using the identity
\begin{equation}
e^{iH^{\dag}t}\eta e^{-iHt}=\eta, \label{eq:id_eta}
\end{equation}
we obtain
\begin{equation}
e^{iHt}\gdualn{\phi}(0,\x)e^{-iHt}=\eta^{-1}\phi^{\dag}_{H}(x)\eta.\label{eq:dphi_t}
\end{equation}
By comparing~(\ref{eq:dphi_t}) with~(\ref{eq:dphi_H_t}), consistency requires  $\gdualn{\phi}_{H}(x)=\eta^{-1}\phi^{\dag}_{H}(x)\eta$. Therefore, both fields $\phi_{H}$ and $\gdualn{\phi}_{H}$ must have the same time evolution. Their product transforms as
\begin{equation}
\gdualn{\phi}_{H}(x)\phi_{H}(x)=e^{iHt}\gdualn{\phi}(0,\x)\phi(0,\x)e^{-iHt}.
\end{equation}
Since $\gdualn{\phi}\phi$ is pseudo Hermitian, using~(\ref{eq:id_eta}), we find
\begin{equation}
e^{iHt}\left[\gdualn{\phi}(0,\x)\phi(0,\x)\right]e^{-iHt}=\eta^{-1}\left[\gdualn{\phi}_{H}(x)\phi_{H}(x)\right]^{\dag}\eta
\end{equation}
which is equivalent to
\begin{equation}
\gdualn{\phi}_{H}(x)\phi_{H}(x)=\left[\gdualn{\phi}_{H}(x)\phi_{H}(x)\right]^{\#}.
\end{equation}

The above analysis of pseudo Hermitian conjugation on scalar fermionic fields also apply to general operators. Given an operator $A$ and its pseudo Hermitian conjugate $A^{\#}$ at $t=0$, they both have the same time evolution in the Heisenberg picture. If an operator is pseudo Hermitian at $t=0$, then it is pseudo Hermitian at all times.

\section{Symplectic symmetry}\label{sym}

Apart from the global $\text{U}(1)$ symmetry, the Lagrangian density also has a global symplectic symmetry $\text{Sp}(2,\mathbb{C})$~\cite{LeClair:2007iy}. Since $\{\phi(x),\gdualn{\phi}(x)\}=0$, the product $\gdualn{\phi}(x)\phi(x)$ can be written as
\begin{equation}
\gdualn{\phi}(x)\phi(x)=\frac{1}{2}\Phi^{\text{T}}(x)\Omega\Phi(x)
\end{equation}
where
\begin{equation}
\Phi(x)=\left[\begin{matrix}
\gdualn{\phi}(x) \\
\phi(x)
\end{matrix}\right],\quad \Omega=\left(\begin{matrix}
0 & 1 \\
-1 & 0
\end{matrix}\right).\label{eq:doublet}
\end{equation}
Therefore, the Lagrangian density is invariant under the symplectic transformations\footnote{If we perform the same analysis for complex bosonic scalar field, we would have obtained the conserved currents and charges associated with the $\text{U}(1)$ symmetry.} 
\begin{equation}
\Phi(x)\rightarrow M\Phi(x),\quad \Phi^{\text{T}}(x)\rightarrow\Phi^{\text{T}}(x)M^{\text{T}}
\end{equation}
where $M$ is a $2\times 2$ complex matrix satisfying
\begin{equation}
M^{\text{T}}\Omega M=\Omega \label{eq:omega}
\end{equation}
with $M^{\text{T}}$ being the transposition of $M$. Equation~(\ref{eq:omega}) is satisfied for all complex matrices of unit determinant. Therefore, they are continuous transformations and can be generated via $M=e^{X}$.
Expand $M$ near the identity
\begin{equation}
\left[I+X^\text{T}+O({X^{\text{T}}}^{2})\right]\Omega\left[I+X+O(X^{2})\right]=\Omega,
\end{equation}
we obtain
\begin{equation}
X^{\text{T}}\Omega=-\Omega X. \label{eq:X}
\end{equation}
Solving~(\ref{eq:X}), we find the general solution
\begin{equation}
X=\left(\begin{matrix}
\theta_{1}+i\theta_{2} & \theta_{3}+i\theta_{4} \\
\theta_{5}+i\theta_{6} & -\theta_{1}-i\theta_{2}
\end{matrix}\right),\quad \theta_{i}\in\mathbb{R}\label{eq:X_soln}
\end{equation}
Substituting~(\ref{eq:X_soln}) into $M$, the generators are given by
\begin{equation}
X_{i}=\frac{dM}{d\theta_{i}}\Big\vert_{\theta_{i}=0}
\end{equation}
from which we obtain
\begin{alignat}{3}
X_{1}&=\left(\begin{matrix}
1 & 0 \\
0 &-1
\end{matrix}\right), &\quad
X_{2}&=\left(\begin{matrix}
i & 0 \\
0 &-i
\end{matrix}\right),&\quad
X_{3}&=\left(\begin{matrix}
0 & 1 \\
0 & 0 
\end{matrix}\right),\\
X_{4}&=\left(\begin{matrix}
0 & i \\
0 & 0
\end{matrix}\right),&\quad
X_{5}&=\left(\begin{matrix}
0 & 0 \\
1 & 0
\end{matrix}\right),&\quad
X_{6}&=\left(\begin{matrix}
0 & 0 \\
i & 0 
\end{matrix}\right).
\end{alignat}
Under the infinitesimal transformation
\begin{equation}
\Phi_{n}\rightarrow\Phi_{n}+\theta_{i}(X_{i})_{nm}\Phi_{m},
\end{equation}
the conserved currents are given by
\begin{equation}
J^{\mu}_{i}=N\left[\frac{\partial\mathscr{L}_{0}}{\partial(\partial_{\mu}\Phi_{n})}(X_{i})_{nm}\Phi_{m}\right].\label{eq:J_mu}
\end{equation}
Due to fermionic statistics, there is an ambiguity in the ordering of $\gdualn{\phi}$ and $\phi$. To deal with this issue, we introduce the operation $N$ to order $\gdualn{\phi}$ to the left of $\phi$. The results are
\begin{align}
J^{\mu}_{1}&=\gdualn{\phi}(\partial^{\mu}\phi)-(\partial^{\mu}\gdualn{\phi})\phi,\label{eq:j1}\\
J^{\mu}_{2}&=i\left[\gdualn{\phi}(\partial^{\mu}\phi)-(\partial^{\mu}\gdualn{\phi})\phi\right],\\
J^{\mu}_{3}&=(\partial^{\mu}\phi)\phi=-\phi(\partial^{\mu}\phi),\\
J^{\mu}_{4}&=i(\partial^{\mu}\phi)\phi,\\
J^{\mu}_{5}&=(\partial^{\mu}\gdualn{\phi})\gdualn{\phi}=-\gdualn{\phi}(\partial^{\mu}\gdualn{\phi}),\\
J^{\mu}_{6}&=i(\partial^{\mu}\gdualn{\phi})\gdualn{\phi}=-i\gdualn{\phi}(\partial^{\mu}\gdualn{\phi}).\label{eq:j6}
\end{align}
To couple the currents to gauge fields, they must be pseudo Hermitian. Given an arbitrary operator $\mathcal{O}$, it can be made pseudo Hermitian by the linear combination $e^{i\vartheta}\mathcal{O}+e^{-i\vartheta}\mathcal{O}^{\#}$ where $\vartheta\in\mathbb{R}$. Applying this procedure to the above currents, we find 
\begin{align}
J^{\mu}_{1}&\rightarrow 2i\sin\vartheta_{1}\left[\gdualn{\phi}(\partial^{\mu}\phi)-(\partial^{\mu}\gdualn{\phi})\phi\right],\\
J^{\mu}_{2}&\rightarrow 2i\cos\vartheta_{2}\left[\gdualn{\phi}(\partial^{\mu}\phi)-(\partial^{\mu}\gdualn{\phi})\phi\right],\\
J^{\mu}_{3}&\rightarrow e^{i\vartheta_{3}}(\partial^{\mu}\phi)\phi+e^{-i\vartheta_{3}}\gdualn{\phi}(\partial^{\mu}\gdualn{\phi}),\\
J^{\mu}_{4}&\rightarrow e^{i\vartheta_{4}}i(\partial^{\mu}\phi)\phi-e^{-i\vartheta_{4}}i\gdualn{\phi}(\partial^{\mu}\gdualn{\phi}),\\
J^{\mu}_{5}&\rightarrow e^{i\vartheta_{5}}(\partial^{\mu}\gdualn{\phi})\gdualn{\phi}+e^{-i\vartheta_{5}}\phi(\partial^{\mu}\phi),\\
J^{\mu}_{6}&\rightarrow e^{i\vartheta_{6}}i(\partial^{\mu}\gdualn{\phi})\gdualn{\phi}-e^{-i\vartheta_{6}}i\phi(\partial^{\mu}\phi),
\end{align}
for $\vartheta_{i}\in\mathbb{R}$. We find the following currents, namely, $J^{\mu}_{1}$, $J^{\mu}_{3}$, and $J^{\mu}_{4}$ to be linearly-dependent on $J^{\mu}_{2}$, $J^{\mu}_{5}$, and $J^{\mu}_{6}$ respectively.
Therefore, there are three linearly-independent currents
\begin{align}
K^{\mu}_{1}&=J^{\mu}_{3}=i\sin\vartheta_{1}\left[\gdualn{\phi}(\partial^{\mu}\phi)-(\partial^{\mu}\gdualn{\phi})\phi\right],\\
K^{\mu}_{2}&=e^{i\vartheta_{3}}(\partial^{\mu}\phi)\phi+e^{-i\vartheta_{3}}\gdualn{\phi}(\partial^{\mu}\gdualn{\phi}),\\
K^{\mu}_{3}&=e^{i\vartheta_{4}}i(\partial^{\mu}\phi)\phi-e^{-i\vartheta_{4}}i\gdualn{\phi}(\partial^{\mu}\gdualn{\phi}).
\end{align}
The corresponding generators are
\begin{equation}
Y_{1}=\sin\vartheta_{1}\left(\begin{matrix}
i & 0 \\
0 &-i
\end{matrix}\right),\quad
Y_{2}=\left(\begin{matrix}
0 & e^{i\vartheta_{3}} \\
-e^{-i\vartheta_{3}} & 0
\end{matrix}\right),\quad
Y_{3}=\left(\begin{matrix}
0 & ie^{i\vartheta_{4}}\\
ie^{-i\vartheta_{4}} &  0
\end{matrix}\right).
\end{equation}
Choosing the phases to be
\begin{equation}
\sin\vartheta_{1}=1,\quad e^{i\vartheta_{3}}=e^{i\vartheta_{4}}=e^{i\vartheta},
\end{equation}
the resulting generators
satisfy the $\text{su}(2)$ algebra
\begin{equation}
[Y_{i},Y_{j}]=2\epsilon_{ijk}Y_{k}.
\end{equation}
The currents become
\begin{align}
K^{\mu}_{1}&=i\left[\gdualn{\phi}(\partial^{\mu}\phi)-(\partial^{\mu}\gdualn{\phi})\phi\right],\\
K^{\mu}_{2}&=e^{i\vartheta}(\partial^{\mu}\phi)\phi+e^{-i\vartheta}\gdualn{\phi}(\partial^{\mu}\gdualn{\phi}),\\
K^{\mu}_{3}&=i\left[e^{i\vartheta}(\partial^{\mu}\phi)\phi-e^{-i\vartheta}\gdualn{\phi}(\partial^{\mu}\gdualn{\phi})\right],
\end{align}
and the conserved charges are given by
\begin{align}
&Q_{1}=i\int d^{3}x\left[\gdualn{\phi}(\partial_{t}\phi)-(\partial_{t}\gdualn{\phi})\phi\right],\label{eq:q1}\\
&Q_{2}=\int d^{3}x\left[e^{i\vartheta}(\partial_{t}\phi)\phi+e^{-i\vartheta}\gdualn{\phi}\partial_{t}\gdualn{\phi}\right],\label{eq:q2}\\
&Q_{3}=i\int d^{3}x\left[e^{i\vartheta}(\partial_{t}\phi)\phi-e^{-i\vartheta}\gdualn{\phi}(\partial_{t}\gdualn{\phi})\right].\label{eq:q3}
\end{align}
Substituting~(\ref{eq:phi}-\ref{eq:dphi}) into~(\ref{eq:q1}-\ref{eq:q3}), we obtain the normal-ordered charges
\begin{align}
Q_{1}&=\int d^{3}p\left[a^{\dag}(\p)a(\p)-b^{\dag}(\p)b(\p)\right],\\
Q_{2}&=i\int d^{3}p\left[e^{-i\vartheta}a^{\dag}(\p)b(\p)+e^{i\vartheta}b^{\dag}(\p)a(\p)\right],\\
Q_{3}&=\int d^{3}p\left[e^{-i\vartheta}a^{\dag}(\p)b(\p)-e^{i\vartheta}b^{\dag}(\p)a(\p)\right].
\end{align}
All three charges are pseudo Hermitian. Additionally, $Q_{1}$ and $iQ_{2,3}$ are Hermitian, satisfying the $\text{su}(2)$ algebra
\begin{align}
[Q_{1},(iQ_{2})]&=2i(iQ_{3}),\\
[(iQ_{2}),(iQ_{3})]&=2iQ_{1},\\
[Q_{1},(iQ_{3})]&=-2i(iQ_{2}).
\end{align}
After normal ordering $Q_{1}$, it defines the charges of the particle and anti-particle, namely,
\begin{equation}
Q_{1}|a\rangle=+|a\rangle,\quad Q_{1}|b\rangle=-|b\rangle.
\end{equation}
For $Q_{2,3}$, they map the particle state to the anti-particle state and vice versa\footnote{LeClair and Neubert asserts that due to the relation $\text{Sp}(2)\cong\text{SO}(3)$, the states are of spin-half. We believe their interpretation to be incorrect. This is because the spin of states are determined by their eigenvalues with respect to one of the Casimir invariant operators of the Poincar\'{e} group, namely $s=-m^{2}j(j+1)$ with $j=0,\frac{1}{2},\cdots$. Since the scalar fermionic fields are constructed from the scalar representation of the Poincar\'{e} group, we have $j=0$ and hence $s=0$.} 
\begin{alignat}{2}
Q_{2}|a\rangle&=ie^{i\vartheta}|b\rangle,\quad & Q_{2}|b\rangle&=ie^{-i\vartheta}|a\rangle,\\
Q_{3}|a\rangle&=-e^{i\vartheta}|b\rangle,\quad & Q_{3}|b\rangle&=e^{-i\vartheta}|a\rangle.
\end{alignat}

By demanding the currents to be pseudo Hermitian, the global $\text{Sp}(2,\mathbf{C})$ symmetry becomes $\text{SU}(2)$. Therefore, the Lagrangian density, including interacting potentials constructed from $\gdualn{\phi}$ and $\phi$ has a global $\text{SU}(2)\times\text{U}(1)$ symmetry. Because gauge group is semi-simple and compact, by treating $\gdualn{\phi}$ and $\phi$ as doublet~(\ref{eq:doublet}), we can couple them to non-Abelian gauge fields that resemble the electroweak sector of the SM. However, we should also note that the doublet structure presented here is fundamentally different from the SM fermionic doublet because it is bosonic and contains only of one specie of particle. It would be interesting to see if the symmetry is preserved in presence of quantum corrections. We leave this task for future investigations.

\section{Generalized unitary evolution}\label{opt}

Interacting Hamiltonians constructed as functions of $\gdualn{\phi}$ and $\phi$ are pseudo Hermitian and hence complex. One may therefore suspect the resulting $S$-matrix to be non-unitary. This concern was partially addressed in~\cite{Ahluwalia:2022zrm}. There, a formalism to compute scattering amplitudes with pseudo Hermitian Hamiltonians and the definition of generalized unitarity relation were proposed. Here we review the formalism and discuss problems to be addressed.

Let $|\alpha\rangle$ and $\langle\beta|$ to be states that evolve under the pseudo Hermitian Hamiltonian
\begin{equation}
|\alpha(t)\rangle=e^{-iHt}|\alpha\rangle, \quad
\langle\beta(t)|=\langle\alpha|e^{iH^{\dag}t}.
\end{equation}
To preserve time translation symmetry, we use the $\eta$-product~\cite{Mostafazadeh:2001jk}
\begin{equation}
\langle\beta|\alpha\rangle_{\eta}\equiv\langle\beta|\eta|\alpha\rangle.
\end{equation}
Using~(\ref{eq:id_eta}), we find $\langle\beta(t)|\alpha(t)\rangle_{\eta}=\langle\beta|\alpha\rangle_{\eta}$. The matrix element of operator must now be defined as
\begin{equation}
A^{(\eta)}_{\beta\alpha}\equiv\langle\beta|\eta A|\alpha\rangle\label{eq:A_ba}
\end{equation}
so that it is invariant under time translation. To see this, we take $U(t)\equiv e^{iHt}$ and find
\begin{align}
A^{(\eta)}_{\beta\alpha}&=\langle\beta|U^{\dag}(t)[U^{\dag-1}(t)\eta U^{-1}(t)][U(t)AU^{-1}(t)]U(t)|\alpha\rangle\nonumber\\
&=\langle\beta(t)|\eta A_{H}(t)|\alpha(t)\rangle\nonumber\\
&=\left[A^{(\eta)}_{H}(t)\right]_{\beta\alpha}.
\end{align}
From~(\ref{eq:A_ba}), we define the expectation value to be $\langle A\rangle_{\eta}\equiv A^{(\eta)}_{\alpha\alpha}$. When $A$ is pseudo Hermitian, the expectation value is real.

In the scattering process $\alpha\rightarrow\beta$ described by pseudo Hermitian Hamiltonian, the $S$-matrix is given by
\begin{equation}
S_{\beta\alpha}\equiv\langle\beta_{+}|\alpha_{-}\rangle_{\eta}. \label{eq:Sba}
\end{equation}
where $|\alpha_{-}\rangle$ and $|\beta_{+}\rangle$ are the 'in' and 'out' states. They are related to the free states by
\begin{align}
|\alpha_{-}\rangle&=\Omega_{-}|\alpha_{0}\rangle,\\
|\beta_{+}\rangle&=\Omega_{+}|\beta_{0}\rangle,
\end{align}
where
\begin{equation}
\Omega(\tau)=e^{iH\tau}e^{-iH_{0}\tau},\quad
\Omega_{\pm}=\lim_{\tau\rightarrow\pm\infty}\Omega(\tau).
\end{equation}
Using $\Omega_{\pm}$, we write~(\ref{eq:Sba}) as
\begin{align}
S_{\beta\alpha}&=\langle\beta_{0}|\Omega^{\dag}_{+}\eta\Omega_{-}|\alpha_{0}\rangle\nonumber\\
&\equiv\langle\beta_{0}|S|\alpha_{0}\rangle 
\end{align}
where $S=\Omega^{\dag}_{+}\eta\Omega_{-}$. Because $H$ is pseudo Hermitian, both $\Omega$ and the $S$-matrix are \textit{non-unitary}. Instead, their inverses are obtained via pseudo Hermitian conjugation
\begin{equation}
\Omega^{-1}=\Omega^{\#},\quad S^{-1}=S^{\#}.
\end{equation}
The matrix component of $S^{-1}$ is given by
\begin{align}
S^{-1}_{\gamma\beta}=S^{\#}_{\gamma\beta}&\equiv\langle\gamma_{0}|S^{\#}|\beta_{0}\rangle\nonumber\\
&=\langle\gamma_{0}|\eta^{-1}S^{\dag}\eta|\beta_{0}\rangle\nonumber\\
&=\langle\gamma_{0}|\Omega^{-1}_{-}\Omega_{+}\eta^{-1}|\beta_{0}\rangle.\label{eq:inv_S}
\end{align}
In obtaining~(\ref{eq:inv_S}), we have used $\eta^{\dag}=\eta$ and $\eta=\eta^{-1}$. Now, we recall that the free states evolve under the free Hamiltonian which is Hermitian. These states admit a Hermitian inner-product that is positive-definite and invariant under time translation
\begin{equation}
\langle\beta_{0}|\alpha_{0}\rangle=\delta(\beta-\alpha).
\end{equation}
Therefore, they satisfy the completeness relation the completeness relation 
\begin{equation}
\int d\beta|\beta_{0}\rangle\langle\beta_{0}|=I \label{eq:free_complete}
\end{equation}
and the $S$-matrix satisfies the identity
\begin{equation}
\int d\beta S^{\#}_{\gamma\beta}S_{\beta\alpha}=\delta(\gamma-\alpha).\label{eq:g_unit}
\end{equation}
This is the \textit{generalized unitarity relation}. For any $S$-matrices defined in terms of the $\eta$-product with pseudo Hermitian Hamiltonians, they will satisfy~(\ref{eq:g_unit}). 

The completeness relation for the free states holds at all times because the free Hamiltonian is Hermitian so that $|\alpha_{0},t\rangle\langle\alpha_{0},t|=|\alpha_{0}\rangle\langle\alpha_{0}|$. As for the in and out states that evolve under pseudo Hermitian Hamiltonians, their completeness relation cannot take the form of~(\ref{eq:free_complete}) since $|\alpha_{\pm},t\rangle\langle\alpha_{\pm},t|\neq|\alpha_{\pm}\rangle\langle\alpha_{\pm}|$. To derive the completeness relation for the in and out states, we use the fact that their $\eta$-product is invariant under time translation and that
\begin{equation}
\langle\beta_{\pm}|\alpha_{\pm}\rangle_{\eta}=\langle\beta_{0}|\alpha_{0}\rangle_{\eta}.
\end{equation}
Using~(\ref{eq:eta}), we find $\eta|\alpha_{0}\rangle=(-1)^{n_{\alpha}}|\alpha_{0}\rangle$ where $n_{\alpha}$ is the total number of anti-particles contained in $|\alpha_{0}\rangle$. Therefore,
\begin{equation}
\langle\beta_{\pm}|\alpha_{\pm}\rangle_{\eta}=(-1)^{n_{\alpha}}\delta(\beta-\alpha).
\end{equation}
The completeness relation then reads
\begin{equation}
\int d\beta (-1)^{n_{\beta}}|\beta_{\pm}\rangle\langle\beta_{\pm}|\eta=I, \label{eq:complete_pm}
\end{equation}
where
\begin{align}
\int d\beta (-1)^{n_{\beta}}|\beta_{\pm}\rangle\langle\beta_{\pm}|\eta|\alpha_{\pm}\rangle&=|\alpha_{\pm}\rangle,
\end{align}
and
\begin{align}
\int d\beta (-1)^{n_{\beta}}\langle\alpha_{\pm}|\eta|\beta_{\pm}\rangle\langle\beta_{\pm}|\eta&=\langle\alpha_{\pm}|\eta.
\end{align}
The completeness relation~(\ref{eq:complete_pm}) holds at all times since $|\alpha_{\pm},t\rangle\langle\alpha_{\pm},t|\eta=|\alpha_{\pm}\rangle\langle\alpha_{\pm}|\eta$. The fact that the $\eta$ product is not positive-definite has important implications for pseudo Hermitian theories to be discussed below and in sec.~\ref{concl}.

The $S$-matrix and its inverse admit the following expansions~\cite{Ahluwalia:2022zrm}
\begin{align}
S_{\beta\alpha}&=\eta_{\beta\alpha}+\sum^{\infty}_{n=1}\frac{(-i)^{n}}{n!}\int d^{4}x_{1}\cdots d^{4}x_{n}\langle\beta_{0}|\eta T\left[\mathscr{H}(x_{1})\cdots\mathscr{H}(x_{n})\right]|\alpha_{0}\rangle,\\
S^{\#}_{\gamma\beta}&=\eta_{\gamma\beta}+\sum^{\infty}_{n=1}\frac{(+i)^{n}}{n!}\int d^{4}x_{1}\cdots d^{4}x_{n}\langle\gamma_{0}|\eta T\left[\mathscr{H}^{\dag}(x_{n})\cdots\mathscr{H}^{\dag}(x_{1})\right]|\beta_{0}\rangle,
\end{align}
where $\eta_{\beta\alpha}=\langle\beta_{0}|\eta|\alpha_{0}\rangle$. Normalizing the $S$-matrix as 
\begin{align}
&S_{\beta\alpha}\equiv \eta_{\beta\alpha}-2\pi iM_{\beta\alpha}\delta^{4}(p_{\beta}-p_{\alpha}),\\
&S^{\#}_{\gamma\beta}\equiv \eta_{\gamma\beta}+2\pi iM^{\#}_{\gamma\beta}\delta^{4}(p_{\gamma}-p_{\beta}),
\end{align}
where $M^{\#}_{\gamma\beta}=\langle\gamma_{0}|\eta^{-1}M^{\dag}\eta|\beta_{0}\rangle$, we obtain the generalized optical theorem
\begin{align}
&i\int d\beta\left[\eta_{\gamma\beta}M_{\beta\alpha}\delta^{4}(p_{\beta}-p_{\alpha})-M^{\#}_{\gamma\beta}\eta_{\beta\alpha}\delta^{4}(p_{\gamma}-p_{\beta})\right]\nonumber\\
&=2\pi\int d\beta
\left[\delta^{4}(p_{\beta}-p_{\gamma})\delta^{4}(p_{\beta}-p_{\alpha})M^{\#}_{\gamma\beta}M_{\beta\alpha}\right]. \label{eq:im}
\end{align}
Setting $\gamma=\alpha$ yields
\begin{align}
i\int d\beta\left(\eta_{\alpha\beta}M_{\beta\alpha}-M^{\#}_{\alpha\beta}\eta_{\beta\alpha}\right)
=2\pi\int d\beta\left[\delta^{4}(p_{\beta}-p_{\alpha})
M^{\#}_{\alpha\beta}M_{\beta\alpha}\right].\label{eq:mm}
\end{align}

The generalized unitarity relation with pseudo Hermitian Hamiltonians is a generalization to unitary quantum mechanics with Hermitian Hamiltonians. For the generalization to be consistent, there must be a prescription to compute transition probabilities. Two criteria are required to ensure consistency. Firstly, the transition probability $P(\alpha\rightarrow\beta)$ for any process $\alpha\rightarrow\beta$ must be positive-definite. Secondly, we need $\sum_{\beta}P(\alpha\rightarrow\beta)=1$. When the $S$-matrix is unitary, both criteria are equivalent and are trivially satisfied. 

The important question is: \textit{How do we compute transition probabilities for pseudo Hermitian theories?} The generalized unitarity relation of the $S$-matrix suggests that the transition probability for $\alpha\rightarrow\beta$ ought to be proportional to $M^{\#}_{\alpha\beta}M_{\beta\alpha}$ but this quantity is not positive-definite. So instead, it was proposed in~\cite{Ahluwalia:2022zrm} that we multiply it by a phase $\wp_{\beta\alpha}$ to obtain $\wp_{\beta\alpha}M^{\#}_{\alpha\beta}M_{\beta\alpha}\geq0$ and interpret this quantity to be proportional to the transition probability. However, this proposal is unsatisfactory for the following reason. If we replace the term $M^{\#}_{\alpha\beta}M_{\beta\alpha}$ in~(\ref{eq:mm}) by~$\wp_{\beta\alpha}M^{\#}_{\alpha\beta}M_{\beta\alpha}$ for processes where $\wp_{\beta\alpha}\neq1$, the generalized optical theorem is no longer satisfied. What this means is that had we adopted this prescription, then we would end up with $\sum_{\beta}P(\alpha\rightarrow\beta)\neq1$ which is unacceptable for any physical theories.

Defining the correct transition probabilities is an important problem to be addressed. In the LN model which involves quartic self-interacting scalar fermions to be studied, the observed difficulty can be ameliorated. To see how it occurs, let us examine $M^{\#}_{\alpha\beta}$. Using the definition of pseudo Hermitian conjugation and the solution of $\eta$ given in~(\ref{eq:eta}), we find
\begin{equation}
M^{\#}_{\alpha\beta}=\langle\alpha_{0}|\eta^{-1}M^{\dag}\eta|\beta_{0}\rangle=e^{-i\pi(n_{\alpha}-n_{\beta})}M^{\dag}_{\alpha\beta}
\end{equation}
where $n_{\alpha}$ and $n_{\beta}$ are the number of anti-particles in states $|\alpha\rangle$ and $|\beta\rangle$ respectively. When $n_{\alpha}=n_{\beta}$ (or up to a integer multiples of $2\pi$), we have $M^{\#}_{\alpha\beta}=M^{\dag}_{\alpha\beta}$. This is equivalent to the demand that the number of anti-particles is a conserved quantity, which is general not satisfied. For the LN model we show that the 2 to 2 scattering process satisfies the generalized unitarity relation. In 2 to 2 scattering, since the number of anti-particles in the initial and final states are the same, the generalized unitarity relation becomes unitary. However, as observed by LeClair, 2 to 4 scattering process (pair-production) becomes possible at high-energy and this spoils unitarity~\cite{LeClair:2024xqo,LeClair:2025qvd} .

\subsubsection*{The LN-model}

We now consider the LN-model~\cite{LeClair:2007iy}
\begin{equation}
\mathscr{L}=\sum^{N}_{i=1}\left(\partial^{\mu}\gdualn{\phi}_{i}\partial_{\mu}\phi_{i}-m^{2}\gdualn{\phi}_{i}\phi_{i}\right)-\frac{g}{2}\left(\sum^{N}_{i=1}\gdualn{\phi}_{i}\phi_{i}\right)^{2}\label{eq:L_phi_N}
\end{equation}
where all the fields have equal mass and anti-commute with each other
\begin{equation}
\left\{\phi_{i}(x),\phi_{j}(x)\right\}=\left\{\phi_{i}(x),\gdualn{\phi}_{j}(x)\right\}=
\left\{\gdualn{\phi}_{i}(x),\gdualn{\phi}_{j}(x)\right\}=0.
\end{equation}
Due to fermionic statistics, $\gdualn{\phi}^{2}_{i}(x)=\phi^{2}_{i}(x)=0$. The interacting density simplifies to\footnote{We can expand the interaction $\int d^{3}x\mathscr{H}$ to show that it is non-Hermitian.}
\begin{equation}
\mathscr{H}=g\sum^{N}_{i<j}\left[\gdualn{\phi}_{i}(x)\phi_{i}(x)\gdualn{\phi}_{j}(x)\phi_{j}(x)\right]. \label{eq:V_s}
\end{equation}
The model is pseudo Hermitian because there exists an $\eta$ given by
\begin{equation}
\eta=\prod^{N}_{i=1}\eta_{i},\quad
\eta_{i}=\exp\left[-i\pi\int d^{3}p\,b^{\dag}_{i}(\p)b_{i}(\p)\right],
\end{equation}
such that $\mathscr{L}^{\#}=\mathscr{L}$, $\mathscr{H}^{\#}=\mathscr{H}$. From~(\ref{eq:V_s}), the following two-body scattering processes are allowed
\begin{align}
ij&\rightarrow i'j',\\
i\bar{j}&\rightarrow i'\bar{j}',\\
i\bar{i}&\rightarrow j'\bar{j'},\quad \text{for all } j\neq i,
\end{align}
where $i,j$ and $\bar{i},\bar{j}$ denote particle and anti-particle states created by the $i$th and $j$th fields respectively. We will now verify that the $S$-matrix for these processes satisfy the generalized unitarity relation up to one-loop.

For all three processes, as there are equal number of particles and anti-particles in the initial and final states, we have $M^{\#}_{\alpha\beta}=M^{\dag}_{\alpha\beta}$. Therefore, the optical theorem~(\ref{eq:mm}) can be rewritten in terms of two-body cross sections
\begin{equation}
i\int d\beta\left(\eta_{\alpha\beta}M_{\beta\alpha}-M^{\dag}_{\alpha\beta}\eta_{\beta\alpha}\right)
=\frac{u_{\alpha}}{8\pi^{3}}\sigma_{\alpha}
\end{equation}
where 
\begin{align}
\sigma_{\alpha}&=\frac{(2\pi)^{4}}{u_{\alpha}}\int d\beta\left[\delta^{4}(p_{\beta}-p_{\alpha})|M_{\beta\alpha}|^{2}\right]\nonumber\\
\end{align}
and
\begin{equation}
u_{\alpha}=\frac{\sqrt{(p_{1}\cdot p_{2})^{2}-m^{4}}}{E_{1}E_{2}}.\label{eq:uij}
\end{equation}
In the center of mass frame $p_{1}=(E,\p)$ and $p_{2}=(E,-\p)$, so that $u_{\alpha}=4|\p|/E_{\text{CM}}$ with $E_{\text{CM}}=2E$. The cross-section simplifies to~\cite[sec.~3.4]{Weinberg:1995mt}
\begin{equation}
\sigma_{\alpha}=\frac{(2\pi)^{4}E^{2}_{\text{CM}}}{16}\int d\Omega|M_{\beta\alpha}|^{2}
\end{equation}
where $d\Omega$ is the differential solid angle of the final particle states. For these processes, the matrix elements of $\eta_{\beta\alpha}$ are given by
\begin{align}
\eta_{(i'j')(ij)}&=\eta_{(ij)(i'j')}=+\delta[(i'j')-(ij)],\\
\eta_{(i'\overline{j}')(i\overline{j})}&=\eta_{(ij)(i'\overline{j}')}=-\delta[(i'\overline{j}')-(i\overline{j})],\\
\eta_{(j'\overline{j}')(i\overline{i})}&=\eta_{(i\overline{i})(j'\overline{j}')}=-\delta[(j'\overline{j}')-(i\overline{i})],
\end{align}
so we obtain
\begin{align}
\text{Im}\left[M_{(ij)(ij)}\right]&=-\frac{1}{8\pi^{3}}\left[1-\frac{4m^{2}}{E^{2}_{\text{CM}}}\right]^{1/2}\sigma_{ij},\label{eq:o1}\\
\text{Im}\left[M_{(i\overline{j})(i\overline{j})}\right]&=+\frac{1}{8\pi^{3}}\left[1-\frac{4m^{2}}{E^{2}_{\text{CM}}}\right]^{1/2}\sigma_{i\overline{j}},\label{eq:o3}\\
\text{Im}\left[M_{(i\overline{i})(i\overline{i})}\right]&=+\frac{1}{8\pi^{3}}\left[1-\frac{4m^{2}}{E^{2}_{\text{CM}}}\right]^{1/2}\sigma_{i\overline{i}},\label{eq:o2}
\end{align}
where we have used $|\p|=\left(1-4m^{2}/E^{2}_{\text{CM}}\right)^{1/2}$.
At tree-level, the cross-sections are given by
\begin{align}
\sigma_{ij}&=\sigma_{i\overline{j}}=\frac{g^{2}}{16\pi E^{2}_{\text{CM}}},\\
\sigma_{i\overline{i}}&=\frac{g^{2}}{16\pi E^{2}_{\text{CM}}}(N-1).
\end{align}
Substituting the cross-sections into the right-hand side of~(\ref{eq:o1}-\ref{eq:o3}) and compare them with the imaginary part of the amplitudes given by~(\ref{eq:i1}-\ref{eq:i3}), we find that the generalized optical theorem is satisfied.

\section{Conclusions} \label{concl}

According to the conventional spin-statistics theorem, scalar fields must furnish bosonic statistics. When phrased as a no-go theorem, it means that anti-commuting scalar fields violate locality and unitarity. The theory of scalar fermion constructed by LN showed that once the Hamiltonian is allowed to be pseudo Hermitian, the no-go theorem no longer applies. We believe this construct to be of fundamental importance because it represents an extension to the spin-statistics theorem. We have shown in this paper, that the theory respects Lorentz and admit generalized unitarity. However, the generalized unitarity relation does not reduce to unitarity.

In our formulation of scattering theory with the indefinite $\eta$ inner-product, unitarity requires the number of anti-particles to be conserved. This is in general, not satisfied. As a result, such theories are usually discarded as being pathological. Nevertheless, for 2 to 2 scattering process below the pair-production threshold, the LN model is unitary, so within our formulation, the model may have applications in condensed matter physics. Alternatively, symplectic fermions has been applied to realized dS/CFT correspondence where unitary evolution may not be essential~\cite{Anninos:2011ui}.



Concerning the quartic self-interaction involving $N$ scalar fermions, this feature is reminiscent of the  Gross-Neveu~\cite{PhysRevD.10.3235} and the Nambu-Jona Lasinio model~\cite{PhysRev.122.345}. Except here, the scalar fermions have renormalizable quartic self-interaction in four space-time dimensions. It would be interesting to study the massless LN model in the large $N$ limit and investigate the possibility of dynamical mass generation.



\acknowledgments

I am grateful to Dharam Vir Ahluwalia for constant discussions and encouragements. I would like to thank James Brister, Ting-Long Feng, Suro Kim, Lei-Hua Liu, Guo-En Nian, Zheng Sun, Han Yan, Wenqi Yu, Cong Zhang and Siyi Zhou discussions. This work supported by The Sichuan University Post-doctoral Research Fund No.~2022SCU12119.


\appendix

%
%

\section{Amplitudes}\label{amp}

We compute the one-loop amplitudes and their imaginary part that are relevant for verifying the generalized optical theorem in sec.~\ref{opt}. Because of the fermionic statistics and the non-trivial adjoint, caution must be exercised when contracting the fields among themselves and with the states. The amplitudes are given by
\begin{align}
S(ij\rightarrow i'j')&=\frac{(-ig)^{2}}{2}\int d^{4}x d^{4}y\langle j'i'|\eta T\left[ (\gdualn{\phi}_{i}\phi_{i}\gdualn{\phi}_{j}\phi_{j})_{x}(\gdualn{\phi}_{i}\phi_{i}\gdualn{\phi}_{j}\phi_{j})_{y}\right]|ij\rangle\nonumber\\
&=\frac{(-ig)^{2}}{2}\int d^{4}x d^{4}y\langle j'i'|T\left[ (\gdualn{\phi}_{i}\phi_{i}\gdualn{\phi}_{j}\phi_{j})_{x}(\gdualn{\phi}_{i}\phi_{i}\gdualn{\phi}_{j}\phi_{j})_{y}\right]|ij\rangle\nonumber\\
&=-\frac{(-ig)^{2}}{2}\int d^{4}x d^{4}y\langle j'i'|T\left[ (\gdualn{\phi}_{i}\gdualn{\phi}_{j}\phi_{i}\phi_{j})_{x}(\gdualn{\phi}_{i}\gdualn{\phi}_{j}\phi_{j}\phi_{i})_{y}\right]|ij\rangle\nonumber\\
&=-(-ig)^{2}\int d^{4}xd^{4}y
\langle
 \wick{
        \c2 {j}' \c1 {i}'
        \vert
        (\gdualn{\c1 \phi}_i \gdualn{\c2\phi}_j}
 \wick{\c2\phi_{i} \c1\phi_{j})_{x}
 (\gdualn{\c2\phi}_{i}\gdualn{\c1\phi}_{j}}
 \wick{\c2 \phi_{j} \c1\phi_{i} )_{y}
        \vert
        \c1 i \c2 j} 
\rangle+\cdots \nonumber\\
&=(-ig)^{2}\int d^{4}xd^{4}y
\langle
 \wick{
        \c2 {j}' \c1 {i}'
        \vert
        (\gdualn{\c1 \phi}_i \gdualn{\c2\phi}_j)_{x}}
 S_{i}(x-y)S_{j}(x-y)
 \wick{(\c2 \phi_{j} \c1\phi_{i} )_{y}
        \vert
        \c1 i \c2 j} 
\rangle+\cdots
\end{align}
\begin{align}
S(i\overline{i}\rightarrow i'\overline{i}')&=\sum^{N}_{j\neq i}\frac{(-ig)^{2}}{2!}\int d^{4}x d^{4}y
\langle \overline{i}'i'|\eta T\left[ (\gdualn{\phi}_{i}\phi_{i}\gdualn{\phi}_{j}\phi_{j})_{x}(\gdualn{\phi}_{i}\phi_{i}\gdualn{\phi}_{j}\phi_{j})_{y}\right]|i\overline{i}\rangle\nonumber\\
&=-\sum^{N}_{j\neq i}\frac{(-ig)^{2}}{2!}\int d^{4}x d^{4}y
\langle \overline{i}'i'|T\left[ (\gdualn{\phi}_{i}\phi_{i}\gdualn{\phi}_{j}\phi_{j})_{x}(\gdualn{\phi}_{i}\phi_{i}\gdualn{\phi}_{j}\phi_{j})_{y}\right]|i\overline{i}\rangle\nonumber\\
&=-\sum^{N}_{j\neq i}\frac{(-ig)^{2}}{2!}\int d^{4}x d^{4}y
\langle \overline{i}'i'|T\left[ (\gdualn{\phi}_{i}\phi_{i}\gdualn{\phi}_{j}\phi_{j})_{x}(\gdualn{\phi}_{j}\phi_{j}\gdualn{\phi}_{i}\phi_{i})_{y}\right]|i\overline{i}\rangle\nonumber\\
&=-(-ig)^{2}\int d^{4}xd^{4}y\sum^{N}_{j\neq i}
\langle
 \wick{
         \c2{\overline{i}'} \c1 i'
        \vert
        (\gdualn{\c1 \phi}_i \c2 \phi_i}
        \wick{\c2{\gdualn{\phi}}_{j}\c1\phi_{j})_{x}(\gdualn{\c1\phi}_{j}\c2\phi_{j}}
        \wick{\c2{\gdualn{\phi}}_{i}\c1 \phi_{i})_{y}|\c1 i \c2{\overline{i}}}\rangle+\cdots\nonumber\\
&=(-ig)^{2}\int d^{4}xd^{4}y\sum^{N}_{j\neq i}
\langle
 \wick{
         \c2{\overline{i}'} \c1 i'
        \vert
        (\gdualn{\c1 \phi}_i \c2 \phi_i}
		S_{j}(x-y)S_{j}(y-x)
        \wick{\c2{\gdualn{\phi}}_{i}\c1 \phi_{i})_{y}|\c1 i \c2{\overline{i}}}\rangle+\cdots\nonumber\\
\end{align}
\begin{align}
S(i\overline{j}\rightarrow i'\overline{j}')&=\frac{(-ig)^{2}}{2!}\int d^{4}x d^{4}y
\langle \overline{j}'i'|\eta T\left[(\gdualn{\phi}_{i}\phi_{i}\gdualn{\phi}_{j}\phi_{j})_{x}(\gdualn{\phi}_{i}\phi_{i}\gdualn{\phi}_{j}\phi_{j})_{y}\right]|i\overline{j}\rangle\nonumber\\
&=-\frac{(-ig)^{2}}{2!}\int d^{4}x d^{4}y
\langle \overline{j}'i'|T\left[(\gdualn{\phi}_{i}\phi_{j}\phi_{i}\gdualn{\phi}_{j})_{x}(\gdualn{\phi}_{i}\phi_{j}\phi_{i}\gdualn{\phi}_{j})_{y}\right]|i\overline{j}\rangle\nonumber\\
&=\frac{(-ig)^{2}}{2!}\int d^{4}x d^{4}y
\langle \overline{j}'i'|T\left[(\gdualn{\phi}_{i}\phi_{j}\phi_{i}\gdualn{\phi}_{j})_{x}(\gdualn{\phi}_{i}\phi_{j}\gdualn{\phi}_{j}\phi_{i})_{y}\right]|i\overline{j}\rangle\nonumber\\
&=(-ig)^{2}\int d^{4}x d^{4}y
\langle\wick{\c2{\overline{j}'}\c1 i'|(\gdualn{\c1\phi}_{i}\c2\phi_{j}}
\wick{\c1\phi_{i}\c2{\gdualn{\phi}}_{j})_{x}
(\gdualn{\c1\phi}_{i}\c2\phi_{j}}\wick{\c2{\gdualn{\phi}}_{j}\c1\phi_{i})_{y}\vert\c1 i \c2{\overline{j}}}\rangle+\cdots\nonumber\\
&=(-ig)^{2}\int d^{4}x d^{4}y\langle\wick{\c2{\overline{j}'}\c1 i'|(\gdualn{\c1\phi}_{i}\c2\phi_{j}}
S_{i}(x-y)S_{j}(y-x)\wick{\c2{\gdualn{\phi}}_{j}\c1\phi_{i})_{y}\vert\c1 i \c2{\overline{j}}}\rangle+\cdots
\end{align}
For the above expressions, we have made explicit the contributions from the $s$-channel as they are the only terms that have non-vanishing imaginary parts. Since all the particles have equal masses, in the center of mass frame, using the expansions of the field and its adjoint given by (\ref{eq:phi}-\ref{eq:dphi}), the contractions of the fields to the initial and final particle states are given by
\begin{align}
\wick{(\c2\phi_{j}\c1\phi_{i})_{y}|\c1 i \c2 j\rangle}&=+e^{-i(p_{i}+p_{j})\cdot y}\left[(2\pi)^{3}E_{\text{CM}}\right]^{-1},\\
\langle\wick{ \c2 j'\c1 i'|(\gdualn{\c1\phi}_{j}\gdualn{\c2\phi}_{i})_{x}}&=+e^{i(p_{i'}+p_{j'})\cdot x}\left[(2\pi)^{3}E_{\text{CM}}\right]^{-1},
\end{align}
\begin{align}
\wick{(\c2{\gdualn{\phi}}_{i}\c1\phi_{i})_{y}|\c1 i \c2{\overline{i}}\rangle}&=-e^{-i(p_{i}+p_{\overline{i}})\cdot y}\left[(2\pi)^{3}E_{\text{CM}}\right]^{-1},\\
\langle
 \wick{
         \c2{\overline{i}'} \c1 i'
        \vert
        (\gdualn{\c1 \phi}_i \c2 \phi_i)_{x}}&=+e^{i(p_{i'}+p_{\overline{i}'})\cdot x}\left[(2\pi)^{3}E_{\text{CM}}\right]^{-1},
\end{align}
\begin{align}
\wick{(\c2{\gdualn{\phi}}_{j}\c1\phi_{i})_{y}\vert\c1 i \c2{\overline{j}}}\rangle&=-e^{-i(p_{i}+p_{\overline{j}})\cdot y}\left[(2\pi)^{3}E_{\text{CM}}\right]^{-1},\\
\wick{\langle\c2{\overline{j}'}\c1 i'|(\gdualn{\c1\phi}_{i}\c2\phi_{j})_{x}}&=+e^{i(p_{i'}+p_{\overline{j}'})\cdot x}\left[(2\pi)^{3}E_{\text{CM}}\right]^{-1}.
\end{align}
In the limit where the initial and final momenta coincide, the $s$-channel amplitudes are given by
\begin{align}
M_{(ij)(ij)}(s)&=+\frac{i}{(2\pi)^{7}E^{2}_{\text{CM}}}\left[-ig(2\pi)^{4}\right]^{2}V(s),\\
M_{(i\overline{i})(i\overline{i})}(s)&=-\frac{i}{(2\pi)^{7}E^{2}_{\text{CM}}}\left[-ig(2\pi)^{4}\right]^{2}(N-1)V(s),\\
M_{(i\overline{j})(i\overline{j})}(s)&=-\frac{i}{(2\pi)^{7}E^{2}_{\text{CM}}}\left[-ig(2\pi)^{4}\right]^{2}V(s),
\end{align}
where
\begin{equation}
V(s)=\left[\frac{i}{(2\pi)^{4}}\right]^{2}\int d^{4}k\left(\frac{1}{k^{2}-m^{2}+i\epsilon}\right)
\left[\frac{1}{(k+p)^{2}-m^{2}+i\epsilon}\right].
\end{equation}
By dimensional regularization, we obtain
\begin{align}
V(s)=-i\pi^{2}\left[\frac{i}{(2\pi)^{4}}\right]^{2}\int^{1}_{0}dx\left\{\ln\left[m^{2}-s(1-x)x\right]+\frac{2}{\epsilon}+\gamma_{E}\right\},
\end{align}
where $\epsilon=(d-4)$ and $\gamma_{E}$ is the Euler constant. The imaginary part of the amplitudes are given by
\begin{align}
\text{Im}\left[M_{(ij)(ij)}\right]&=-\frac{g^{2}}{128\pi^{4}E^{2}_{\text{CM}}}\left[1-\frac{4m^{2}}{E^{2}_{\text{CM}}}\right]^{1/2},\label{eq:i1}\\
\text{Im}\left[M_{(i\overline{i})(i\overline{i})}\right]&=+\frac{g^{2}}{128\pi^{4}E^{2}_{\text{CM}}}(N-1)\left[1-\frac{4m^{2}}{E^{2}_{\text{CM}}}\right]^{1/2},\label{eq:i2}\\
\text{Im}\left[M_{(i\overline{j})(i\overline{j})}\right]&=+\frac{g^{2}}{128\pi^{4}E^{2}_{\text{CM}}}\left[1-\frac{4m^{2}}{E^{2}_{\text{CM}}}\right]^{1/2}.\label{eq:i3}
\end{align}

\bibliography{Bibliography}

\providecommand{\href}[2]{#2}\begingroup\raggedright\begin{thebibliography}{10}

\bibitem{Duck:1997ua}
I.~Duck and E.~C.~G. Sudarshan, {\em {Pauli and the spin-statistics theorem}}.
\newblock 1997.

\bibitem{Streater:1989vi}
R.~F. Streater and A.~S. Wightman, {\em {PCT, spin and statistics, and all
  that}}.
\newblock 1989.

\bibitem{Weinberg:1995mt}
S.~Weinberg, {\em {The Quantum theory of fields. Vol. 1: Foundations}}.
\newblock Cambridge University Press, 6, 2005.

\bibitem{Planck:2018vyg}
{\bf Planck} Collaboration, N.~Aghanim et~al., {\it {Planck 2018 results. VI.
  Cosmological parameters}},  {\em Astron. Astrophys.} {\bf 641} (2020) A6,
  [\href{http://xxx.lanl.gov/abs/1807.0620}{{\tt arXiv:1807.0620}}]. [Erratum:
  Astron.Astrophys. 652, C4 (2021)].

\bibitem{Dirac:1931kp}
P.~A.~M. Dirac, {\it {Quantised singularities in the electromagnetic field,}},
  {\em Proc. Roy. Soc. Lond. A} {\bf 133} (1931), no.~821 60--72.

\bibitem{Bender:1998ke}
C.~M. Bender and S.~Boettcher, {\it {Real spectra in nonHermitian Hamiltonians
  having PT symmetry}},  {\em Phys. Rev. Lett.} {\bf 80} (1998) 5243--5246,
  [\href{http://xxx.lanl.gov/abs/physics/9712001}{{\tt physics/9712001}}].

\bibitem{Bender:2007nj}
C.~M. Bender, {\it {Making sense of non-Hermitian Hamiltonians}},  {\em Rept.
  Prog. Phys.} {\bf 70} (2007) 947,
  [\href{http://xxx.lanl.gov/abs/hep-th/0703096}{{\tt hep-th/0703096}}].

\bibitem{Mostafazadeh:2001jk}
A.~Mostafazadeh, {\it {PseudoHermiticity versus PT symmetry. The necessary
  condition for the reality of the spectrum}},  {\em J. Math. Phys.} {\bf 43}
  (2002) 205--214, [\href{http://xxx.lanl.gov/abs/math-ph/0107001}{{\tt
  math-ph/0107001}}].

\bibitem{Mostafazadeh:2008pw}
A.~Mostafazadeh, {\it {Pseudo-Hermitian Representation of Quantum Mechanics}},
  {\em Int. J. Geom. Meth. Mod. Phys.} {\bf 7} (2010) 1191--1306,
  [\href{http://xxx.lanl.gov/abs/0810.5643}{{\tt arXiv:0810.5643}}].

\bibitem{RevModPhys.15.175}
W.~Pauli, {\it On dirac's new method of field quantization},  {\em Rev. Mod.
  Phys.} {\bf 15} (Jul, 1943) 175--207.

\bibitem{LeClair:2007iy}
A.~LeClair and M.~Neubert, {\it {Semi-Lorentz invariance, unitarity, and
  critical exponents of symplectic fermion models}},  {\em JHEP} {\bf 10}
  (2007) 027, [\href{http://xxx.lanl.gov/abs/0705.4657}{{\tt
  arXiv:0705.4657}}].

\bibitem{Ahluwalia:2022ttu}
D.~V. Ahluwalia, J.~M.~H. da~Silva, C.-Y. Lee, Y.-X. Liu, S.~H. Pereira, and
  M.~M. Sorkhi, {\it {Mass dimension one fermions: Constructing darkness}},
  {\em Phys. Rept.} {\bf 967} (2022) 1--43,
  [\href{http://xxx.lanl.gov/abs/2205.0475}{{\tt arXiv:2205.0475}}].

\bibitem{Ahluwalia:2022zrm}
D.~V. Ahluwalia and C.-Y. Lee, {\it {Spin-half bosons with mass dimension
  three-half: Evading the spin-statistics theorem}},  {\em EPL} {\bf 140}
  (2022), no.~2 24001, [\href{http://xxx.lanl.gov/abs/2212.0945}{{\tt
  arXiv:2212.0945}}]. [Erratum: EPL 140, 69901 (2022)].

\bibitem{Ahluwalia:2022yvk}
D.~V. Ahluwalia, J.~M.~H. da~Silva, and C.-Y. Lee, {\it {Mass dimension one
  fields with Wigner degeneracy: A theory of dark matter}},  {\em Nucl. Phys.
  B} {\bf 987} (2023) 116092, [\href{http://xxx.lanl.gov/abs/2212.1311}{{\tt
  arXiv:2212.1311}}].

\bibitem{Ryu:2022ffg}
S.~Ryu and J.~Yoon, {\it {Unitarity of Symplectic Fermion in $\alpha$-vacua
  with Negative Central Charge}},
  \href{http://xxx.lanl.gov/abs/2208.1216}{{\tt arXiv:2208.1216}}.

\bibitem{Anninos:2011ui}
D.~Anninos, T.~Hartman, and A.~Strominger, {\it {Higher Spin Realization of the
  dS/CFT Correspondence}},  {\em Class. Quant. Grav.} {\bf 34} (2017), no.~1
  015009, [\href{http://xxx.lanl.gov/abs/1108.5735}{{\tt arXiv:1108.5735}}].

\bibitem{Ng:2012xp}
G.~S. Ng and A.~Strominger, {\it {State/Operator Correspondence in Higher-Spin
  dS/CFT}},  {\em Class. Quant. Grav.} {\bf 30} (2013) 104002,
  [\href{http://xxx.lanl.gov/abs/1204.1057}{{\tt arXiv:1204.1057}}].

\bibitem{Das:2012dt}
D.~Das, S.~R. Das, A.~Jevicki, and Q.~Ye, {\it {Bi-local Construction of
  Sp(2N)/dS Higher Spin Correspondence}},  {\em JHEP} {\bf 01} (2013) 107,
  [\href{http://xxx.lanl.gov/abs/1205.5776}{{\tt arXiv:1205.5776}}].

\bibitem{Sato:2015tta}
Y.~Sato, {\it {Comments on Entanglement Entropy in the dS/CFT Correspondence}},
   {\em Phys. Rev. D} {\bf 91} (2015), no.~8 086009,
  [\href{http://xxx.lanl.gov/abs/1501.0490}{{\tt arXiv:1501.0490}}].

\bibitem{Narayan:2016xwq}
K.~Narayan, {\it {On $dS_4$ extremal surfaces and entanglement entropy in some
  ghost CFTs}},  {\em Phys. Rev. D} {\bf 94} (2016), no.~4 046001,
  [\href{http://xxx.lanl.gov/abs/1602.0650}{{\tt arXiv:1602.0650}}].

\bibitem{Fei:2015kta}
L.~Fei, S.~Giombi, I.~R. Klebanov, and G.~Tarnopolsky, {\it {Critical Sp(N )
  models in 6 \ensuremath{-} \ensuremath{\epsilon} dimensions and higher spin
  dS/CFT}},  {\em JHEP} {\bf 09} (2015) 076,
  [\href{http://xxx.lanl.gov/abs/1502.0727}{{\tt arXiv:1502.0727}}].

\bibitem{Robinson:2009xm}
D.~J. Robinson, E.~Kapit, and A.~LeClair, {\it {Lorentz Symmetric Quantum Field
  Theory for Symplectic Fermions}},  {\em J. Math. Phys.} {\bf 50} (2009)
  112301, [\href{http://xxx.lanl.gov/abs/0903.2399}{{\tt arXiv:0903.2399}}].

\bibitem{LeClair:2024xqo}
A.~LeClair, {\it {A rich structure of renormalization group flows for
  Higgs-like models in 4 dimensions}},
  \href{http://xxx.lanl.gov/abs/2411.0747}{{\tt arXiv:2411.0747}}.

\bibitem{LeClair:2025qvd}
A.~LeClair, {\it {Non-perturbative renormalization group for pseudo-Hermitian
  scalar fields in 4D}},  {\em J. Phys. A} {\bf 59} (2026), no.~16 165206,
  [\href{http://xxx.lanl.gov/abs/2504.0932}{{\tt arXiv:2504.0932}}].

\bibitem{PhysRevD.10.3235}
D.~J. Gross and A.~Neveu, {\it Dynamical symmetry breaking in asymptotically
  free field theories},  {\em Phys. Rev. D} {\bf 10} (Nov, 1974) 3235--3253.

\bibitem{PhysRev.122.345}
Y.~Nambu and G.~Jona-Lasinio, {\it Dynamical model of elementary particles
  based on an analogy with superconductivity. i},  {\em Phys. Rev.} {\bf 122}
  (Apr, 1961) 345--358.

\end{thebibliography}\endgroup
\bibliographystyle{JHEP}


\end{document}